\documentclass[prb,aps,letter,superscriptaddress,showpacs,twocolumn,floatfix]{revtex4}

\usepackage{graphicx}
\usepackage[ansinew]{inputenc}
\usepackage{array}
\usepackage{color}
\usepackage{amsmath}
\usepackage{amsxtra}
\usepackage{amstext}
\usepackage{amssymb}
\usepackage{latexsym}
\usepackage{dsfont}
\begin{document}

\title{Spatially resolved spectroscopy of monolayer graphene on SiO$_2$}

\author{A.	Deshpande}
\affiliation{Department of Physics, University of Arizona, Tucson, AZ, 85721 USA.}
\author{W. Bao}
\author{F. Miao}
\author{C.N. Lau}
\affiliation{Department of Physics, University of California at Riverside, Riverside, CA 92521 USA.}
\author{B.J. LeRoy}
\email{leroy@physics.arizona.edu}
\affiliation{Department of Physics, University of Arizona, Tucson, AZ, 85721 USA.}

\date{\today}

\begin{abstract}
We have carried out scanning tunneling spectroscopy measurements on exfoliated monolayer graphene on SiO$_2$ to probe the correlation between its electronic and structural properties.  Maps of the local density of states are characterized by electron and hole puddles that arise due to long range intravalley scattering from intrinsic ripples in graphene and random charged impurities. At low energy, we observe short range intervalley scattering which we attribute to lattice defects. Our results demonstrate that the electronic properties of graphene are influenced by intrinsic ripples, defects and the underlying SiO$_2$ substrate.
\end{abstract}


\maketitle
Graphene is the two dimensional form of carbon characterized by a honeycomb lattice with two inequivalent lattice sites. It is unusual in many aspects, one of them being the linear energy dispersion relation that causes the electrons to obey the relativistic Dirac equation instead of the Schrödinger equation \cite{novoselov2007}. The theoretical prediction of the instability of 2D crystals along with unsuccessful efforts at their synthesis on insulating substrates kept graphene far from the realm of experiments. A breakthrough in the isolation of graphene from 3D graphite opened up a new frontier for experimental investigation \cite{novoselov2004,novoselovPNAS}. Electrical transport measurements have brought to the fore some of the exotic properties of graphene like a novel quantum Hall effect, high carrier mobility and minimum conductivity \cite{novoselov2005,zhang2005,miao2007}. Transmission electron microscopy characterization (TEM) of graphene revealed intrinsic ripples \cite{meyer2007} which were confirmed by Monte Carlo simulations \cite{fasolino2007}. These structural findings hinted that electron scattering and carrier localization could be highly influenced not only by external effects such as impurities but also by intrinsic ripples. Furthermore, electrical transport measurements showed the suppression of electron localization \cite{morozov2006} and that the mobility was insensitive to doping \cite{schedin2007}. Both these results, unexpected for a two dimensional system like graphene, were attributed to corrugations in graphene.  Hence, to investigate phenomena involving electronic and structural aspects of graphene we have performed spatially resolved spectroscopy measurements.

A single atom thick graphene sheet can be studied by suspending it on a micron-sized metal grid or by supporting it on a substrate like SiO$_2$ \cite{meyer2007,ishigami2007,stolyarova2007}. The sheet is not perfectly flat but contoured with intrinsic ripples with deformations up to 1 nm normal to the plane clearly evident from TEM of suspended graphene and Monte Carlo simulations \cite{meyer2007,fasolino2007}.  These ripples are due to the stability requirement of the two dimensional lattice and the ability of carbon to bond with asymmetric bond lengths \cite{meyer2007,fasolino2007}.  Supported sheets of graphene have an additional constraint on their morphology due to the SiO$_2$ substrate. They have been found to partially conform to the substrate while still having intrinsic ripples and therefore areas not in contact with SiO$_2$ \cite{ishigami2007,stolyarova2007,geringer2009}. These previous measurements showed the topographic structure of monolayer graphene. Spectroscopy measurements of graphene on SiO$_2$ revealed the contribution of phonons to the tunneling process \cite{zhang2008}. However, none of these measurements probed the effects of graphene morphology on its electronic properties. Theoretical calculations show that variations in the local curvature of graphene, ripples, lead to a $\pi-\sigma$ orbital mixing which changes the local electrochemical potential breaking particle hole symmetry and causing charge inhomogeneities \cite{kim2008}. The shift in the local chemical potential $\Phi(r)$ is proportional to the square of the curvature of graphene
\begin{equation}
\Phi(r)= -\alpha\frac{3a^{2}}{4}(\nabla^2(h(r))^2
\end{equation}
where $h(r)$ is the local height referenced to a flat configuration, $a$ is the nearest neighbor distance and $\alpha \approx$ 9.2 eV is a constant \cite{kim2008}. This implies that curvature of the graphene sheet leads to density variations due to the shifting chemical potential.

In addition to the intrinsic rippling in graphene, its two dimensional nature makes its entire surface area susceptible to adsorption and defects. Given the intricacies involved in graphene device preparation the presence of random impurities, adsorbates or defects from various sources is difficult to prevent.  Depending on the type of imperfection, they can either act as long or short range scattering sites.  Coulomb scattering by random charged impurities at the graphene-SiO$_2$ boundary is one example of long range scattering. This long range, intravalley, scattering tends to create inhomogeneities in the carrier density forming electron and hole puddles \cite{hwang2007,martin2008}. When the scattering potential has components shorter than the lattice constant, typically due to lattice defects, intervalley scattering takes place which mixes the two sublattices of graphene \cite{ando}. Here we use a low temperature scanning tunneling microscope (STM) to probe the morphology, local electronic properties and scattering phenomena in monolayer graphene.

Graphene was prepared by mechanical exfoliation of graphite on SiO$_2$ \cite{novoselov2004,novoselov2005}. Monolayer areas were identified using an optical microscope and then Ti/Au electrodes were deposited using standard electron beam lithography. The lithography process leaves some PMMA resist on the surface of graphene.  To eliminate the resist the device was annealed in argon and hydrogen at 400 °C for 1 hour \cite{meyer2007} followed by annealing in air at 300 °C for 30 minutes. The device was then immediately transferred to the STM (Omicron low temperature STM operating at T = 4.5 K in ultrahigh vacuum (p $\leq 10^{-11}$ mbar)).  Electrochemically etched tungsten tips were used for imaging and spectroscopy.  All of the tips used were first checked on an Au surface to ensure that their density of states was constant.

\begin{figure}[]
\includegraphics[width=0.48 \textwidth]{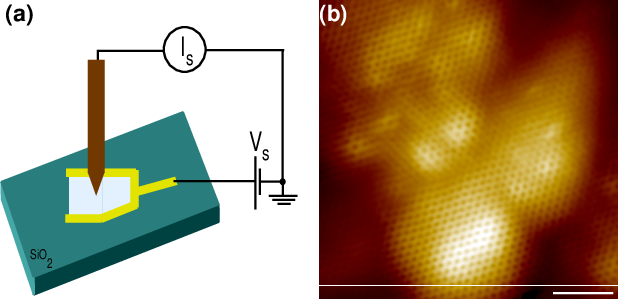}
\caption{(a) Schematic of a mechanically exfoliated monolayer graphene flake (gray) with gold electrodes on a SiO$_2$ substrate. (b) STM topographic image of monolayer graphene showing the hexagonal lattice as well as the underlying surface corrugations. The scale bar is 2 nm. The imaging parameters are sample voltage $V_s$ = 0.25 V, tunneling current $I_s$ = 100 pA.} \label{schematic}
\end{figure}

A schematic of the measurement set-up showing the graphene flake on SiO$_2$ with a gold electrode for electrical contact is shown in Fig. \ref{schematic}(a).  A typical STM image of the monolayer graphene showing the complete hexagonal lattice along with surface corrugations due to the underlying SiO$_2$ substrate is shown in Fig. \ref{schematic}(b). The observation of both sublattices, giving the hexagonal structure, is characteristic of monolayer graphene as opposed to bulk graphite where only a single sublattice is usually observed giving a triangular pattern. The hexagonal lattice in our images extends over areas as large as 40 nm demonstrating the cleanliness of the surface.  The uneven surface underneath the hexagons is a distinctive feature of all our images. This additional corrugation has a height variation of $\sim$ 5Å over an area of 30 $\times$ 30 nm$^2$.  The lateral extent of these corrugations is in the range of a few nanometers mimicking the SiO$_2$ corrugation \cite{ishigami2007,stolyarova2007}.

\begin{figure}[]
\includegraphics[width=0.48 \textwidth]{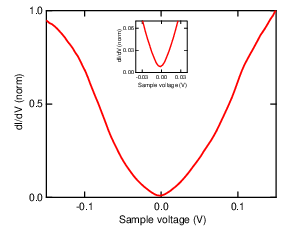}
\caption{dI/dV point spectroscopy shows a linear relationship between the tunneling conductance over a sample voltage $V_s$ of $\pm$ 0.15 V.  The inset shows the region near the Fermi level, no gap is seen.} \label{pointspec}
\end{figure}

We have recorded the local density of states of graphene using dI/dV point spectroscopy measurements. In this case, the tip is fixed at a specific location on the sample, the feedback loop is turned off and the sample voltage is ramped within a specific energy window. An ac modulation voltage of 4 mV rms, 574 Hz is applied to the sample and the resulting dI/dV spectrum is recorded using lockin detection. One such spectrum is shown in Fig. \ref{pointspec}. The dI/dV spectrum is approximately linear in energy and featureless. Neither a gap nor a dip is seen around the Fermi level.

\begin{figure}[]
\includegraphics[width=0.48 \textwidth]{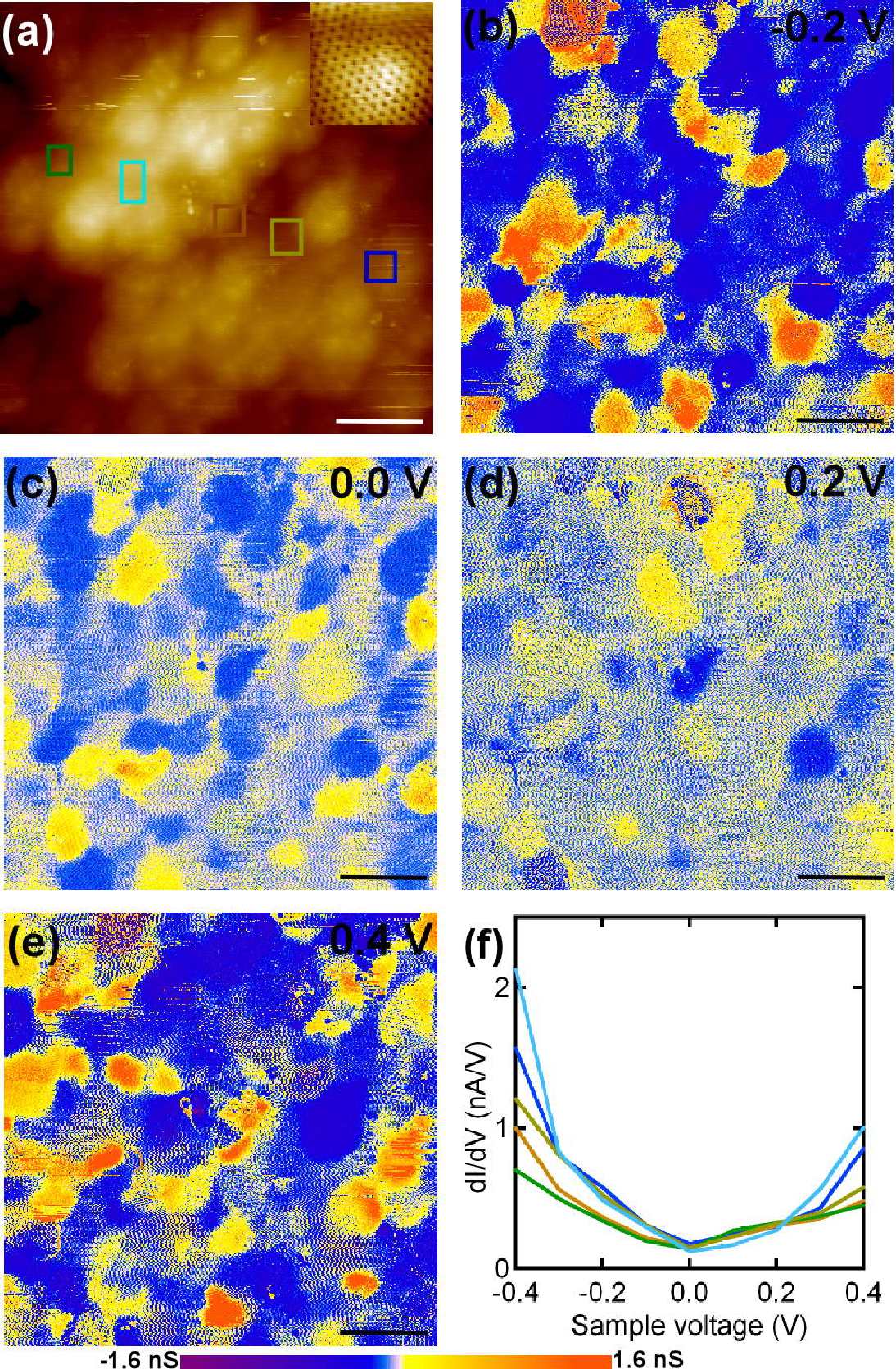}
\caption{(a) STM topography showing the lattice and surface corrugations. The scale bar is 8 nm.  The topography is recorded with $V_s$ = 0.4 V and $I_s$ = 100 pA.  (b) to (e) dI/dV maps at sample voltages -0.2 V, 0.0 V, 0.2 V and 0.4 V respectively showing electron and hole puddles.  For all of the images, the current was stabilized at the parameters of (a) and the feedback was then switched off.  The maps were recorded using lock-in detection. (f) Spatially averaged dI/dV curves at 5 different regions of the graphene indicated in (a). The curves show a crossover near 0.2 V as a result of the shifting Dirac point.} \label{puddles}
\end{figure}

To understand the effect of the corrugations seen in Fig. \ref{schematic}(b) on the density of states of graphene, we have performed spatially resolved scanning tunneling spectroscopy (STS) measurements.  By measuring the differential conductance, dI/dV as a function of energy and position, we obtain maps of the local density of states (LDOS).  Figure \ref{puddles} shows atomically resolved topography and spectroscopy over a 40 $\times$ 40 nm$^2$ area. The topography of the area, Fig. \ref{puddles}(a), demonstrates the atomic resolution as well as the corrugations.  Fig. \ref{puddles}(b)-(e) are maps of the LDOS over the same area for four different sample voltages ranging from -0.2 V to 0.4 V.  The maps show areas of high and low differential conductance which change as a function of energy.  These changes demonstrate that the Dirac point varies as a function of position as discussed below.  At low energy, the shifting of the Dirac point leads to electron and hole puddles.

\begin{figure}[]
\includegraphics[width=0.48 \textwidth]{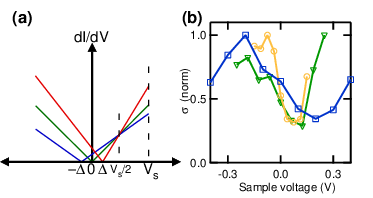}
\caption{(a)Illustration of effecting of shifting Dirac point on dI/dV curves.  The shift in the Dirac point causes the slopes of the dI/dV curves to change leading to a crossover at $V_s/2$. Furthermore, the value of dI/dV at $V_s$ gives information about the amount of the shift in the Dirac point, $\Delta$.  (b) Standard deviation of dI/dV for 3 different locations and values of $V_s$.  The setpoint voltage $V_s$ is 0.4 V (blue curve), 0.25 V (green curve) and 0.15 V (yellow curve). The minimum occurs at $V_s/2$ for each of the curves. } \label{cross}
\end{figure}

When performing the spectroscopy measurements, the feedback circuit is turned off and the tip height is held constant.  The height is determined by the setpoint current, $I_s$, (100 pA in this case) and the setpoint voltage, $V_s$, (0.4 V in this case).  If we assume that the Dirac point is at the Fermi energy, then the dI/dV curve can be written as $\frac{dI}{dV}={\alpha}{|V|}$ where $V$ is the sample voltage (green curve in Fig. \ref{cross}). The value of $\alpha$ is determined by the setpoint parameters, $I_s$ and $V_s$.  This is because the feedback circuit adjusts the tip height such that the integrated current from 0 to $V_s$ is $I_s$. The value of $\alpha$ is determined by setting the area under the dI/dV curve equal to $I_s$, $I_s=\frac{\alpha V_s^2}{2}$. This gives a value of $\alpha=\frac{2I_s}{V_s^2}$ and therefore $\frac{dI}{dV}=\frac{2I_s}{V_s^2}|V|$ (green curve in Fig. \ref{cross}(a)).

However, in a region where the Dirac point shifts by an amount $\Delta<0$ the sample essentially becomes more conductive.  Therefore, the tip must move farther from the surface to maintain the setpoint current causing the dI/dV curve to have a lower slope.  Now, $\frac{dI}{dV}=\beta|V-\Delta|$ as shown by the blue curve in Fig. \ref{cross}(a).  The value of $\beta$ is once again determined by the requirement that the area under the $\frac{dI}{dV}$ curve from 0 to $V_s$ is $I_s$.  This gives $\beta=\frac{2I_s}{V_s^2-2\Delta V_s}$.  The point where two dI/dV curves with slopes $\alpha$ and $\beta$ cross can be obtained by setting the dI/dV curves to be equal and solving for V. The result is that the two curves intersect at $V_s/2$.

When the Dirac point shifts by an amount $\Delta>0$, $\frac{dI}{dV}=\gamma|V-\Delta|$ and a similar effect occurs except the tip moves closer to the surface and the slope of the dI/dV curve increases (red curve in Fig. \ref{cross}(a)). The value of $\gamma$ is given by
$$\gamma=\frac{2I_s}{V_s^2-2\Delta V_s+ 2\Delta^2}$$
Again the dI/dV curves cross at $V_s /2$ if  $\Delta << V_s$ and at negative sample voltages, the curves have different slopes based on the energy of the Dirac point but they do not cross.

This analysis makes two predictions for the LDOS maps. (1) The variation in the maps should be smallest at $V_s/2$ because regardless of the energy of the Dirac point, all the dI/dV curves have the same value at $V_s/2$.  (2) The value of dI/dV at $V_s$ is related to the energy of the Dirac point by the relationship $\Delta=V_s(\frac{1-\eta/2}{1-\eta})$, where $\eta$ is the normalized differential conductance given by $\eta=\frac{dI/dV}{I/V}$.

We have studied the variation in the LDOS maps for a set of three different locations and tunneling parameters.  The results are shown in Fig. \ref{cross}(b).  The blue curve is for the images shown in Fig. \ref{puddles} where $V_s$ was 0.4 V.  From the curve it is clear that the minimum variation occurs at 0.2 V as can also be seen in Fig. \ref{puddles}(d).  The green curve is taken in a different region of the sample with $V_s= 0.25$ V.  The minimum has shifted to lower energy compared to the blue curve and it is now at 0.125 V.  Lastly, the yellow curve has $V_s = 0.15$ V and the minimum is at the lowest energy of the three curves. This shows that the location of the minimum in the LDOS maps is dependent on the parameters used for the measurement but the value of dI/dV at $V_s$ still gives information on the Dirac point.

Fig. \ref{puddles}(f) shows the dI/dV curves taken from the LDOS maps for five different areas of the image as shown in Fig. \ref{puddles}(a).  These curves exhibit a range of dI/dV values at $V_s$ which is evidence for the formation of electron and hole puddles.  In Fig. \ref{puddles}(e), the red regions (large dI/dV) are locations where the Dirac point has shifted towards positive sample voltage while in the blue regions (smaller dI/dV) the Dirac point has shifted towards negative sample voltage.  From the changes in the dI/dV curves we estimate that the shift in the Dirac point,  $\Delta E_d$, is about 77 mV in our images.

\begin{figure}[]
\includegraphics[width=0.48 \textwidth]{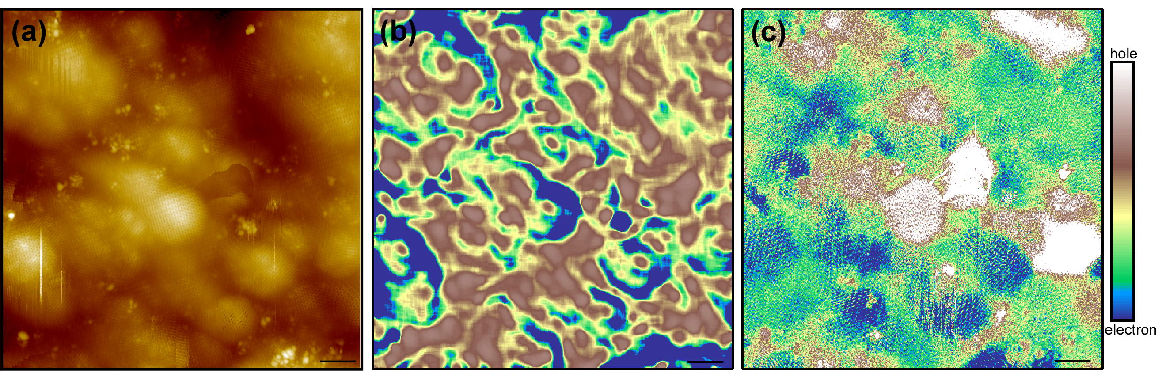}
\caption{(a) STM topographic image of 30x30 nm$^2$ area taken with $V_s$ = 0.15 V and $I_s$ = 50 pA. (b) Shift in local electrochemical potential calculated from the curvature of the topographic image shown in (a). (c) LDOS map recorded at $V_s$ = 0.15 V with the tip height determined using the imaging parameters from the topography measurement in (a). Scale bar in all images is 3 nm.} \label{curve}
\end{figure}

As discussed earlier in this manuscript intrinsic ripples in graphene give rise to a spatially varying electrochemical potential (equation (1)). This equation predicts that highly curved regions of the sample will be electron doped while flat regions will be hole doped. Using a STM topographic image, Fig. \ref{curve}(a), we have calculated the shift in the electrochemical potential caused by the graphene curvature using equation (1). The results in Fig. \ref{curve}(b) show areas of high curvature, large negative change in chemical potential, in blue while relatively flat regions are brown.  A comparison of the shift in Dirac point from the dI/dV map at 0.15 V, Fig. \ref{curve}(c) and the electrochemical potential landscape Fig. \ref{curve}(b) shows there are limited regions which are in agreement between them. For example the curved region in the bottom right side is electron doped (blue region in Fig. \ref{curve}(c) and Fig. \ref{curve}(b)).  However, this is not always the case; there are curved regions such as the blue patch in the center of Fig. \ref{curve}(b) that is hole doped as seen in the center of Fig. \ref{curve}(c). Thus from our measurements we conclude that the curvature in the graphene flake contributes to a variation in the electrochemical potential but it is not the main factor responsible for the features in the dI/dV map. Instead, the potential variation is due to a combination of the ripples and long range scatterers such as random charged impurities present on the graphene sheet \cite{martin2008}. These electron and hole puddles, a signature of disorder in graphene, are also responsible for the finite minimum conductivity at the Dirac point \cite{hwang2007}.

A second source of potential variation in graphene is long range scattering from random charged impurities present at the graphene substrate interface. From our measured shift in the Dirac point, $\Delta E_d$, we calculate the density variation using $\Delta n =\frac{(\Delta E_d)^2}{\pi \hbar^2 v_F^2}$ to be $4.3\times 10^{11}$cm$^{-2}$. An independent measurement of the impurity density can be done using the voltage dependence of the conductivity, $\sigma=20 e \epsilon V_g / (h n_i t)$ where $\epsilon$ is the dielectric constant of SiO$_2$, $t$ is the oxide thickness and $V_g$ is the gate voltage \cite{adam2007}.  We measured another flake of graphene processed in the same manner as the one used for the STS measurements and found $n_i = 4\times 10^{11}$cm$^{-2}$. This impurity density can be converted to local electron density fluctuations using $\delta n^2 = n_i / (8\pi d^2)$ where $d$ is the distance of the impurities from the graphene\cite{hwang2007}.  This agreement between our spectroscopy measurements and transport measurements as well as transport measurements by other groups \cite{hwang2007} indicates that the role of charged impurities is critical for understanding current graphene devices.

\begin{figure}[]
\includegraphics[width=0.48 \textwidth]{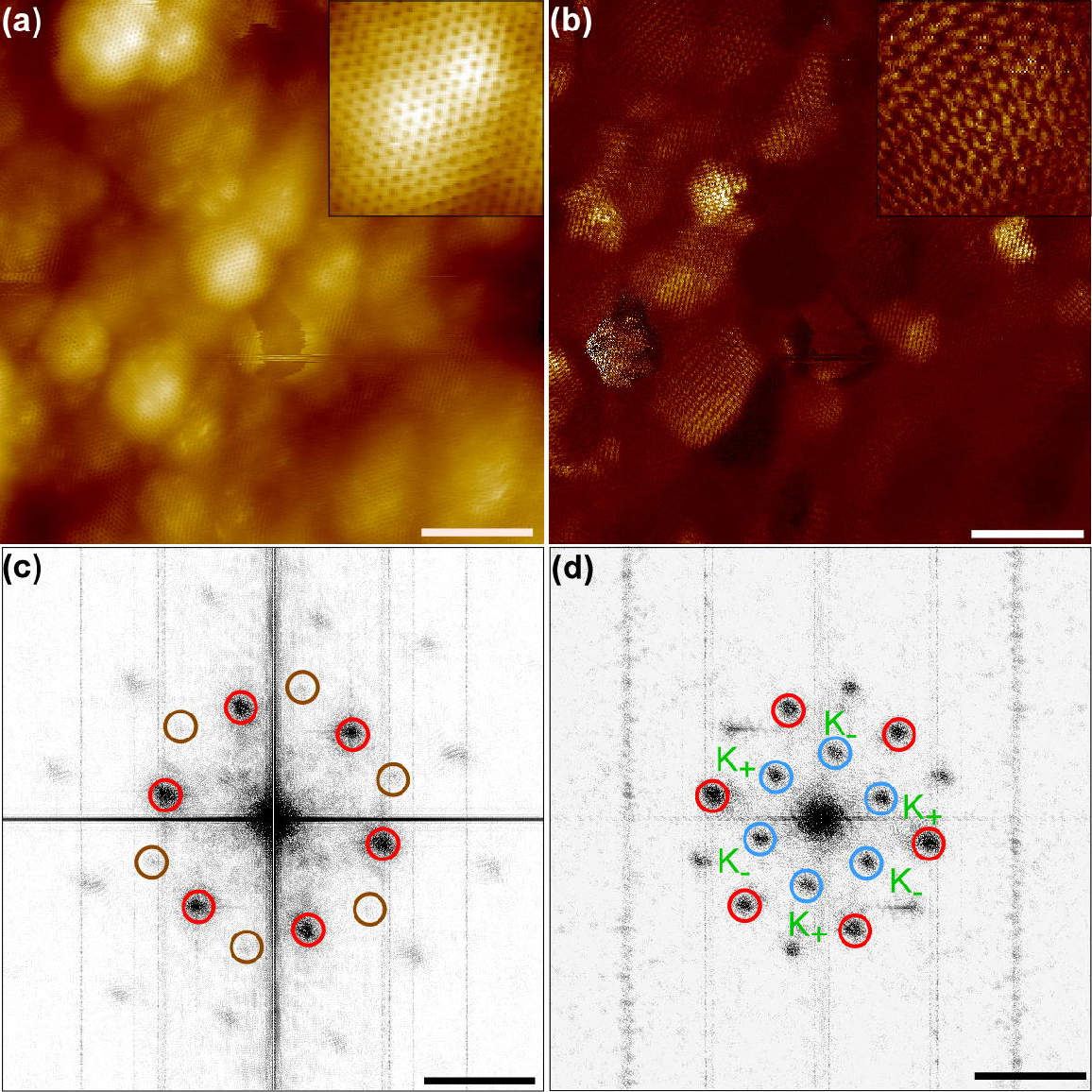}
\caption{(a) STM image of graphene, the scale bar is 6 nm. Inset is a 4 nm region within the image showing the hexagonal lattice. (b) LDOS map at the Fermi energy (0 V) taken simultaneously.  Inset corresponds to the same 4 $\times$ 4 nm$^{2}$ area as in the topography image showing a different pattern due to scattering from defects.  (c) 2D Fourier transform of (a) showing the reciprocal lattice (red circles) and components due to the C-C bonds (brown circles). The scale bar is 4 nm$^{-1}$ . (d) Fourier transform of (b). The inner hexagon (blue circles) represents the interference pattern due to intervalley scattering. The outer hexagon (red circles) is due to the reciprocal lattice.} \label{transforms}
\end{figure}

While long range scatterers give rise to intravalley scattering, within one sublattice, creating the electron and hole puddles seen in the LDOS images, short range scatterers such as lattice defects can also be present in graphene. These short range scatterers induce intervalley scattering from one Dirac cone to the other. The mobility of graphene was found to be insensitive to doping with gas molecules, long range scatterers \cite{schedin2007}.  This implies that there is another source which may already limit the mobility of the graphene such as short range scatterers which tend to give a constant resistivity and hence low mobility \cite{katsnelson2008,chen2008}. We observe lattice defects in the monolayer graphene images and analyze the resulting Fourier transforms of the LDOS maps and topographs for an insight into intervalley scattering. In the case of epitaxial graphene and graphite, lattice defects have been shown to give rise to scattering and interference \cite{rutter2007,ruffieux2005}. Figure \ref{transforms} shows the results for graphene on SiO$_2$. Atomically resolved topography, Fig. \ref{transforms}(a), shows the hexagonal lattice of graphene and the corresponding Fourier transform, Fig. \ref{transforms}(c), contains the reciprocal lattice points (red circles) and components due to the C-C bonds (brown circles) which are longer and rotated by 30 degrees.  Figure \ref{transforms}(b) is an image of the LDOS at the Fermi level showing the defect induced interference pattern superimposed on the electron and hole puddles due to the long range scattering.  The Fourier transform of this image, Fig. \ref{transforms}(d), shows two hexagonal patterns.  The outer hexagon, red circles, is due to the reciprocal lattice and is located at the same points as in Fig. \ref{transforms}(c).  There is also an additional inner hexagon, blue circles, that is rotated by 30 degrees due to scattering.  This is the $\sqrt{3} \times \sqrt{3}$ R30$^\circ$ interference pattern. The six peaks at the K$_\pm$ points in Fig. \ref{transforms}(d), are a result of intervalley, short range, scattering events.

\begin{figure}[]
\includegraphics[width=0.48 \textwidth]{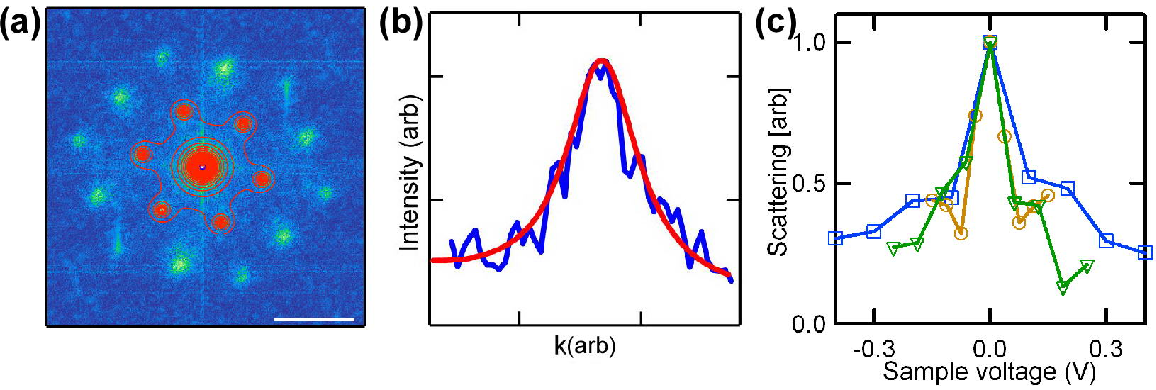}
\caption{(a) 2D Fourier transform of a LDOS map recorded at the Fermi energy. The traces represent the contours corresponding to a 2D Lorentzian fit. The scale bar is 3 nm$^{-1}$. (b) Linescan across one of the peaks in the Fourier transform (blue trace) and the corresponding Lorentzian fit (red trace).   (c) Amount of intervalley scattering as a function of sample voltage. Each curve is from a different region of the graphene flake. The blue curve corresponds to the area shown in Fig. \ref{puddles} and is taken with $V_s$ = 0.4 V.  The green curve and the yellow curve are from different (30 $\times$ 30 nm$^2$) areas taken with $V_s$ = 0.25 V and 0.15 V respectively.} \label{scattering}
\end{figure}

To characterize the strength of intervalley scattering as a function of energy, we have fit the six peaks at the K$_\pm$ points of the $\sqrt{3} \times \sqrt{3}$ R30$^\circ$ interference pattern with a 2D Lorentzian to determine the area under the peak. We do not observe any structures around these K$_\pm$ points unlike those observed in case of graphene on SiC \cite{brihuega2008}. Hence we have chosen a Lorentzian for the fit. This area is then divided by the average value of the Fourier transform to get the relative amount of intervalley scattering at a given energy.  As an example, Figure \ref{scattering}(a) shows one such Fourier transform with six peaks that are fit with a 2D Lorentzian. Figure \ref{scattering}(b) shows the linescan across one of the 6 peaks in the Fourier transform and the Lorentzian fit corresponding to that peak. Figure \ref{scattering}(c) plots the relative strength of these peaks as a function of sample voltage for three different regions of the sample. The broad peak in scattering at the Fermi energy is independent of the imaging parameters. It is consistently observed in areas of the flake with scatterers or defects. As we move away from the Fermi energy the strength of scattering decreases sharply. This enhanced intervalley scattering at low energy is evidence of weak localization of carriers \cite{rutter2007,morgenstern2002}.

In conclusion, we have presented an extensive topographic and spectroscopic investigation of monolayer graphene on SiO$_2$ using a STM at 4.5 K.  We were able to atomically resolve large areas of exfoliated monolayer graphene, record energy resolved local density of states maps and interpret the maps to be a signature of the shifting Dirac point. Also we identified and analyzed intervalley scattering mechanisms on exfoliated monolayer graphene. Thus, the electronic properties of monolayer graphene are closely related to intrinsic ripples, SiO$_2$ substrate morphology and random impurities. Investigation of monolayer graphene flakes can be extended to different geometries and substrates to advance our understanding of the subtleties of the electronic properties of graphene.

CNL, WB and FM acknowledge the support of NSF CAREER DMR/0748910, NSF CBET/0756359 and ONR/DMEA Award H94003-07-2-0703.



\begin{thebibliography}{10}

\bibitem{novoselov2007}
K.S.  Novoselov and A.K. Geim, Nat. Mater. {\bf 6}, 183 (2007).

\bibitem{novoselovPNAS}
K.S. Novoselov \textit{et al.}, Proc. Natl. Acad. Sci. USA {\bf 102}, 10451 (2005).

\bibitem{novoselov2004}
K.S. Novoselov \textit{et al.}, Science {\bf 306}, 666 (2004).

\bibitem{novoselov2005}
K.S. Novoselov \textit{et al.}, Nature (London) {\bf 438}, 197 (2005).

\bibitem{zhang2005}
Y.B. Zhang \textit{et al.}, Nature (London) {\bf 438}, 201 (2006).

\bibitem{miao2007}
F. Miao \textit{et al.}, Science {\bf 317}, 1530 (2007).

\bibitem{meyer2007}
J.C. Meyer \textit{et al.}, Nature (London) {\bf 446}, 60 (2007).

\bibitem{fasolino2007}
A. Fasolino, J.H. Los, and M.I. Katsnelson, Nature Mater. {\bf 6}, 858 (2007).

\bibitem{morozov2006}
S.V. Morozov \textit{et. al},.Phys. Rev. Lett. {\bf 97}, 016801 (2006).

\bibitem{schedin2007}
F. Schedin \textit{et. al}, Nature Mater. {\bf 6}, 652 (2007).

\bibitem{ishigami2007}
M.J. Ishigami \textit{et al.} Nano Lett. {\bf 7}, 1643 (2007).

\bibitem{stolyarova2007}
E. Stolyarova \textit{et al.} Proc. Natl. Acad. Sci. USA {\bf 104}, 9209 (2007).

\bibitem{geringer2009}
V. Geringer \textit{et. al}, Phys. Rev. Lett. {\bf 102}, 076102 (2009).

\bibitem{zhang2008}
Y. Zhang \textit{et. al}, Nature Phys. {\bf 4}, 627 (2008).

\bibitem{kim2008}
E.A. Kim and A.H. Castro Neto, EuroPhys. Lett. {\bf 84}, 57007 (2008).

\bibitem{hwang2007}
E.H. Hwang, S. Adam, and S. Das Sarma, Phys. Rev. Lett. {\bf 98}, 186806 (2007).

\bibitem{martin2008}
J. Martin \textit{et. al}, Nature Phys. {\bf 4}, 144 (2008).

\bibitem{ando}
T. Ando and T. Nakanishi, J. Phys. Soc. Jpn {\bf 67}, 1704 (1998); T. Ando, Phil. Trans. R. Soc. A {\bf 366}, 221 (2008).

\bibitem{adam2007}
S. Adam \textit{et al.}, Proc. Natl. Acad. Sci. U.S.A. {\bf 104}, 18392 (2007).

\bibitem{katsnelson2008}
M.I. Katsnelson and A.K. Geim, Phil. Trans. R. Soc. A {\bf 366}, 195 (2008).

\bibitem{chen2008}
J.H. Chen \textit{et al.}, Nature Phys. {\bf 4}, 377 (2008).

\bibitem{rutter2007}
G.M. Rutter \textit{et al.}, Science {\bf 317}, 219 (2007).

\bibitem{ruffieux2005}
P. Ruffieux \textit{et al.}, Phys. Rev. B {\bf 71}, 153403 (2005).

\bibitem{brihuega2008}
I. Brihuega \textit{et al.}, Phys. Rev. Lett. {\bf 101}, 206802 (2008).

\bibitem{morgenstern2002}
M. Morgenstern \textit{et al.}, Phys. Rev. Lett. {\bf 89}, 136806 (2002).

\end{thebibliography}
\end{document}